\documentclass[12pt]{article}%
\usepackage{ amsmath, graphics, rotating}%

\begin{document}
\large
\begin{center}{\large\bf ELEMENTARY PARTICLES IN A QUANTUM 
THEORY OVER A GALOIS FIELD}\end{center}
\vskip 1em \begin{center} {\large Felix M. Lev} \end{center}
\vskip 1em \begin{center} {\it Artwork Conversion Software Inc.,
1201 Morningside Drive, Manhattan Beach, CA 90266, USA
(Email:  felixlev314@gmail.com)} \end{center}
\vskip 1em

{\it Abstract:}
\vskip 0.5em

We consider elementary particles in a quantum theory based on a Galois field. 
In this approach infinities cannot exist, the cosmological constant problem does not arise 
and one irreducible representation of the symmetry algebra necessarily describes a particle
and its antiparticle simultaneously. {\it In other words, the very existence of antiparticles
is a strong indication that nature is described rather by a finite field (or at least a field with
a nonzero characteristic) than by complex numbers.} As a consequence, the spin-statistics theorem 
is simply a requirement that standard quantum theory should be based on complex numbers and 
elementary particles cannot be 
neutral.  The Dirac vacuum energy problem has a natural solution and the vacuum energy 
(which in the standard theory is infinite and negative) equals zero as it should be.  

\begin{flushleft} PACS: 02.10.De, 03.65.Ta, 11.30.Fs, 11.30.Ly\end{flushleft}

\begin{flushleft} Keywords: quantum theory, Galois fields, elementary particles\end{flushleft}

\section{The statement of the problem}
\label{S1}

The problem of infinities is probably the most
challenging one in standard formulation of
quantum theory. As noted by Weinberg \cite{Wein},
{\it 'Disappointingly this problem appeared with even
greater severity in the early days of quantum theory,
and although greatly ameliorated by
subsequent improvements in the theory, it remains
with us to the present day'}. A desire
to have a theory without divergencies is probably the
main motivation for developing modern theories
extending standard local quantum field theory (QFT),
e.g. loop quantum gravity, noncommutative quantum 
theory, string theory etc.

There exists a wide literature aiming to solve
the difficulties of QFT by replacing the field
of complex numbers by quaternions, p-adic numbers
or other constructions. On the other hand, it is
obvious that the most radical approach is to
consider quantum theory over a Galois field (GFQT). 
Indeed, since any Galois field is finite, 
the problem of infinities in GFQT does not exist 
in principle and all operators are well defined. 

We believe that GFQT is more natural than standard
quantum theory and some arguments are given below
(see Ref. \cite{hep} for a more detailed discussion).

Standard mathematics proceeds from axioms, which are accepted 
without proof. For example, we cannot verify that $a+b=b+a$ for
any numbers $a$ and $b$. At the same time, in the spirit of
quantum theory only those statements should be treated as physical,
which can be experimentally verified, at least in principle.
Suppose we wish to verify that 100+200=200+100. In the
spirit of quantum theory it is insufficient to just say
that 100+200=300 and 200+100=300. We should describe an experiment
where these relations can be verified. In particular,
we should specify whether we have enough resources to
represent the numbers 100, 200 and 300. We believe the
following observation is very important: although standard
mathematics is a part of our everyday life, people typically
do not realize that {\it standard mathematics is implicitly 
based on the assumption that one can have any desirable 
amount of resources}.

\begin{sloppypar}
Suppose, however that our Universe is finite. Then the
amount of resources cannot be infinite and
it is natural to assume that there exists a
fundamental number $p$ such that all calculations can
be performed only modulo $p$. In that case it is natural
to consider a quantum theory over a Galois field with
the characteristic $p$. Since any Galois field is finite, 
the fact that arithmetic in this field is correct can be
verified (at least in principle) by using a finite amount of
resources.
\end{sloppypar}

Note also that standard division obviously reflects our
everyday experience that any macroscopic object can be 
divided by 2, 10, 1000 etc. However, in view of the
existence of elementary particles, the notion of 
division has only a limited applicability. Indeed, we cannot divide by two the
electron or neutrino. This might be an indication that, in the spirit of Ref. 
\cite{Planat}, the ultimate quantum theory 
will be based even not on a Galois field but on a Galois ring
(this observation was pointed out to me by Metod Saniga).
However, in the present paper we will consider a case of 
Galois fields.

If one accepts the idea to replace complex numbers
by a Galois field, the problem arises what
formulation of the standard quantum theory is
most convenient for that purpose. A well known 
historical fact is that originally
quantum theory has been proposed in two formalisms
which seemed essentially different: the
Schroedinger wave formalism and the Heisenberg
operator (matrix) formalism. It has been shown later
by Born, von Neumann and others that the both
formalisms are equivalent and, in addition, the
path integral formalism has been developed. 

In the spirit of the wave or path integral approach
one might try to replace classical spacetime 
by a finite lattice which may even be not a field.
In that case the problem arises what is the natural
'quantum of spacetime' and some physical quantities 
should necessarily have the field structure. A 
detailed discussion can be found in Ref.
\cite{Galois} and references therein. In contrast
to those approaches, we propose to generalize
the standard operator formulation, where quantum 
systems are described by elements of a projective
complex Hilbert spaces and physical quantities
are represented by self-adjoint operators in
such spaces. Recall that in the operator formalism one may
use spacetime and Lagrangians
but their role is only auxiliary: to construct 
proper Hilbert spaces and operators, since this is
all we need to have the maximum possible information
in quantum theory.

In view of the above discussion, GFQT could be 
defined as a theory where
\begin{itemize}
\item {\it Quantum states are represented by 
elements of a 
linear projective space over a Galois field 
and physical quantities are represented by 
linear operators in that space.}
\end{itemize}

The next question is what Galois field should
be used in GFQT. It is well known (see e.g. standard textbooks 
\cite{VDW}) that any Galois
field can contain only $p^n$ elements where $p$
is prime and $n$ is natural. Moreover, the 
numbers $p$ and $n$ define the Galois field
up to isomorphism. 

It is natural to require that there should exist a correspondence
between any new theory and the old one, i.e. at some conditions the both theories should
give close predictions. In particular, there should exist a large number of quantum states
for which the probabilistic interpretation is valid. Then, as shown in  
our papers \cite{lev2,hep}, the number $p$ should necessarily be very large and we
have to understand whether there exist
deep reasons for choosing a particular
value of $p$ or this is simply an accident
that our Universe has been created with this
value. In any case, if we accept that $p$ is
a universal constant then the problem arises
what the value of $n$ is. Since there should
exist the correspondence between GFQT and the 
complex version of standard quantum theory, a 
natural idea is to accept that the principal
number field in GFQT is the Galois field analog
of complex numbers which can be constructed as
follows.

Let $F_p=Z/pZ$ be the residue field modulo $p$
and $F_{p^2}$ be a set of $p^2$ elements $a+bi$ 
where $a,b\in F_p$ and $i$ is
a formal element such
that $i^2=-1$. The question arises whether $F_{p^2}$ is a
field, i.e.
one can define all the arithmetic operations excepting
division by zero.
The definition of addition, subtraction and multiplication
in $F_{p^2}$
is obvious and, by analogy with the field of complex
numbers, one could define division as
$1/(a+bi)\,=a/(a^2+b^2)\,-ib/(a^2+b^2)$ if $a$ and
$b$ are not equal to zero simultaneously.
This definition can be meaningful only if
$a^2+b^2\neq 0$ in $F_p$. If $a$ and $b$ are not
simultaneously equal to zero, this condition can
obviously be reformulated such that $-1$ should not
be a square in $F_p$ (or in terminology of number
theory it should not be a quadratic residue).
We will not consider the case $p=2$ and therefore $p$
is necessarily odd.
Then we have two possibilities: the value of $p\,(mod \,4)$ 
is either 1 or 3. The well known result of number theory
(see e.g. the
textbooks \cite{VDW}) is that -1 is a quadratic residue 
only in the former case and a quadratic nonresidue
in the latter one. Therefore the above
construction of
the field $F_{p^2}$ is correct only if
$p=3\,\,(mod \,4)$.

As shown in Refs. \cite{lev2,hep}, if the principal number
field in GFQT is $F_{p^2}$ where $p$ is very large and
$p=3\,\,(mod \,4)$, there indeed exists the 
correspondence between GFQT and the standard theory.
However, since we treat GFQT as a more general theory
than the standard one, it is desirable not to postulate
that GFQT is based on $F_{p^2}$ (with $p=3\,\,(mod \,4)$)
because standard theory is based on complex numbers
but vice versa, explain the fact that standard 
theory is based on complex numbers since GFQT is based 
on $F_{p^2}$. Therefore we should find a motivation for
the choice of $F_{p^2}$ in GFQT. A possible motivation 
is discussed in Ref. \cite{complex} and another 
motivation is given in the present paper. 

Let us now discuss how one should define the 
notion of elementary particles. Although particles 
are observables and fields are not,
in the spirit of QFT, fields are more fundamental
than particles, and a possible definition is as
follows \cite{Wein1}: {\it It is simply a particle whose
field appears in the Lagrangian. It does not matter if
it's stable, unstable, heavy, light --- if its field
appears in the Lagrangian then it's elementary,
otherwise it's composite.} Another approach has been 
developed by Wigner in his investigations of unitary 
irreducible representations (IRs) of the Poincare 
group \cite{Wigner}. In view of this approach, one 
might postulate that a particle is elementary if the 
set of its wave functions is the space of a unitary IR
of the symmetry group or Lie algebra in the given theory.
In the standard theory the Lie algebras are usually
real and one considers their representations in
complex Hilbert spaces. 

In view of these remarks, it is natural to
define the elementary particle in GFQT as follows.
Let ${\cal A}$ be a Lie algebra over $F_p$ which 
is treated as a symmetry algebra. Then the particle is
elementary if the set of its states forms an IR 
of ${\cal A}$
in $F(p^n)$ (see Ref. \cite{hep} for a detailed 
discussion). 

Representations of Lie algebras in 
spaces with nonzero characteristic are called 
modular representations. There exists a wide theory
of such representations. One of the well known result
is the Zassenhaus theorem \cite{Zass} that any modular 
IR is finite dimensional. In the present paper we 
do not need the general theory since all the results
are obtained explicitly. 

As shown in Refs. \cite{lev2,hep}, standard theories based on
de Sitter (dS) algebra so(1,4) or anti de Sitter (AdS) algebra 
so(2,3) can be generalized to theories based on a Galois field
while the theory based on Poincare algebra cannot. The reasons are as follows.
It is clear that in theories based on Galois fields there can be no
dimensional quantities and all physical quantities are discrete.
In standard dS or AdS invariant theories all physical quantities are dimensionless and 
discrete in units $\hbar/2=c=1$. 
From the formal point of view, the representation operators of the
Poincare algebra also can be chosen dimensionless, e.g. in the Planck
units. However, representations of the Poincare algebra over Galois
fields cannot describe a reasonable physics since in the standard theory the
momentum operators have the continuous spectrum. Therefore the requirement
that a theory should have a physically reasonable generalization to the
case of Galois fields, excludes Poincare invariance as an exact symmetry.

The recent
astronomical data (see e.g. Ref. \cite{Perlmutter}) indicate that the
cosmological constant is small and positive. This is an argument in
favor of so(1,4) vs. so(2,3). On the other hand, in QFT and its
generalizations (string theory, M-theory etc.) a theory based on so(1,4)
encounters serious difficulties and the choice of so(2,3) is preferable
(see e.g. Ref. \cite{Witten}). IRs of the so(2,3) algebra have much in common
with IRs of Poincare algebra. In particular, in IRs of the so(2,3)
algebra the AdS Hamiltonian is either strictly positive or strictly negative
and the supersymmetric generalization is possible. For these reasons in the present paper
for illustration of what happens when complex numbers are replaced by a Galois 
field we assume that ${\cal A}$ is the
modular analog of the algebra so(2,3). 

Since in standard theory the operators 
representing physical quantities are Hermitian,
the question arises what is the analog of this
property in GFQT. Consider first the case 
when $n=2$ and $p=3\,(mod \,4)$. Then the
only nontrivial automorphism of $F_{p^2}$
coincides with complex conjugation 
$z\rightarrow {\bar z}$. One can define the 
scalar product $(...,...)$ in the
representation space such that 
$(x,y)\in F_{p^2}$ and
\begin{equation}
(x,y) =\overline{(y,x)},\quad (ax,y)=\bar{a}(x,y),\quad
(x,ay)=a(x,y)
\label{1}
\end{equation}
Then the operator $B$ is called adjoint to $A$ if 
$(Bx,y)=(y,Ax)$ for any $x,y$ from the 
representation space. As usual,
in this case one can write $A^*=B$. If $A^*=A$, the
operator is called self-adjoint or Hermitian (in
finite dimensional spaces these properties are 
equivalent). 

In the general case the above scalar 
product does not define a positive definite
metric but there exists a large amount of
states for which probabilistic interpretation
is possible. The main idea of
establishing the correspondence between
GFQT and the standard theory is as follows
(see Ref. \cite{hep} for a
detailed discussion). One can denote the
elements of $F_p$ not only as $0,1,...p-1$
but also as $0,\pm 1,...,\pm (p-1)/2$.
Such elements of $F_p$ are called minimal
residues \cite{VDW}. Let $f$ be a map from
$F_p$ to $Z$ such that $f(a)$ has the same
notation in $Z$ as its minimal residue in
$F_p$. Then for elements $a,b$ such that 
$|f(a)|,|f(b)|\ll p$, addition, subtraction 
and multiplication in $F_p$ and $Z$ are
the same, i.e. $f(a\pm b)=f(a)\pm f(b)$
and $f(ab)=f(a)f(b)$. In other words, $f$
is a local homomorphism of the rings $F_p$
and $Z$.  

It is well known \cite{VDW} that the field 
$F_{p^n}$ has $n-1$ nontrivial
automorphisms. Therefore, if $n$ is arbitrary,
a scalar product and Hermiticity can be defined
in different ways. We do not
assume from the beginning that $n=2$ and 
$p=3\,\,(mod \,4)$. Our results do not 
depend on the explicit choice
of the scalar product and ${\bar z}$
is used to denote an element obtained from
$z\in F_{p^n}$ by the automorphism compatible 
with the scalar product in question.

The paper is organized as follows. In Sects.
\ref{S2} and \ref{S3} we construct
modular IRs describing elementary 
particles in GFQT and their quantization
is discussed in Sects. \ref{S4} and \ref{S5}.
The main 
results of the paper --- the solution of
the Dirac vacuum energy problem in GFQT and
spin-statistics theorem --- are discussed in
Sects. \ref{S6} and \ref{S7}, respectively.
Although some results require extensive
calculations, they involve only finite sums
in Galois fields. For this reason all the 
results can be reproduced even by readers
who previously did not have practice in 
calculations with Galois fields. A more 
detailed description of calculations can
be found in Ref. \cite{hep}.

\section{Modular IRs of the sp(2) algebra}
\label{S2}

The key role in constructing modular IRs of 
the so(2,3)
algebra is played by modular IRs of the sp(2) 
subalgebra. They are described by a set of 
operators $(a',a",h)$ satisfying the 
commutation relations
\begin{equation}
[h,a']=-2a'\quad [h,a"]=2a"\quad [a',a"]=h
\label{2}
\end{equation}
The  Casimir operator of the second order for 
the algebra (\ref{2}) has the form
\begin{equation}
K=h^2-2h-4a"a'=h^2+2h-4a'a"
\label{3}
\end{equation}

We first consider representations with the 
vector $e_0$ such that
\begin{equation}
a'e_0=0,\quad he_0=q_0e_0
\label{4}
\end{equation}
where $q_0\in F_p$, $f(q_0)\ll p$ and $f(q_0) > 0$.
Recall that we consider the representation in a
linear space over $F_{p^k}$ where $k$ is a natural
number (see the discussion in Sect. \ref{S1}).
Denote $e_n =(a")^ne_0$.
Then it follows from Eqs. (\ref{3}) and (\ref{4}),
that \begin{equation}
he_n=(q_0+2n)e_n,\quad Ke_n=q_0(q_0-2)e_n,
\label{5}
\end{equation}
\begin{equation}
a'a"e_n=(n+1)(q_0+n)e_n
\label{6}
\end{equation}

One can consider analogous representations in the
standard theory. Then $q_0$ is a positive real
number, $n=0,1,2,...$ and the elements $e_n$ form 
a basis of the IR. In this case $e_0$ is a vector 
with a minimum eigenvalue of the
operator $h$ (minimum weight) and there are no
vectors with the maximum weight.
The operator $h$ is positive definite
and bounded below by the quantity $q_0$. For these
reasons the above modular IRs can be treated as
modular analogs of such standard IRs that $h$ is
positive definite.

Analogously, one can construct modular IRs 
starting from
the element $e_0'$ such that
\begin{equation}
a"e_0'=0,\quad he_0'=-q_0e_0'
\label{7}
\end{equation}
and the elements $e_n'$ can be defined as
$e_n'=(a')^ne_0'$.
Such modular IRs are analogs of standard 
IRs where $h$ is negative definite.
However, in the modular case Eqs. (\ref{4})
and (\ref{7}) define the same IRs. This is
clear from the following consideration.

The set $(e_0,e_1,...e_N)$ will be a basis of
IR if $a"e_i\neq 0$
for $i<N$ and $a"e_N=0$. These conditions
must be compatible with $a'a"e_N=0$.
Therefore, as follows from
Eq. (\ref{6}), $N$ is defined by the condition 
$q_0+N=0$ in $F_p$. As a result, if
$q_0$ is one of the numbers $1,...p-1$
then $N=p-q_0$ and the dimension of IR is equal
to $p-q_0+1$ (in agreement with the Zassenhaus
theorem \cite{Zass}). It is easy to see that
$e_N$ satisfies Eq. (\ref{7}) and therefore
it can be identified with $e_0'$.

Let us forget for a moment that the eigenvalues of the
operator $h$ belong to $F_p$ and will treat them as
integers. Then, as follows from Eq. (\ref{5}), the
eigenvalues are
$$q_0,q_0+2,...,2p-2-q_0, 2p-q_0.$$
Therefore, if $f(q_0)>0$ and $f(q_0)\ll p$,
the maximum value of $q_0$ is equal to
$2p-q_0$, i.e. it is of order $2p$. 

\section{Modular IRs of the so(2,3) algebra}
\label{S3}

Standard IRs of the AdS so(2,3) algebra relevant for 
describing elementary particles have been 
considered by many authors. The description
in this section is a combination of two elegant ones
given in Ref. \cite{Evans} for standard IRs and Ref.
\cite{Braden} for modular IRs. 
In the standard theory the representation
operators of the so(2,3) algebra in units
$\hbar/2=c=1$ are given by 
\begin{equation}
[M^{ab},M^{cd}]=-2i (g^{ac}M^{bd}+g^{bd}M^{cd}-
g^{ad}M^{bc}-g^{bc}M^{ad})
\label{8}
\end{equation}
where $a,b,c,d$ take the values 0,1,2,3,5 and the operators
$M^{ab}$ are antisymmetric. The diagonal metric tensor
has the
components $g^{00}=g^{55}=-g^{11}=-g^{22}=-g^{33}=1$. 
In these units the spin of fermions is odd, and
the spin of bosons is even. If $s$ is the particle spin
then the corresponding IR of the su(2) algebra has
the dimension $s+1$. 

Note that our definition of the AdS symmetry on
quantum level does not involve the cosmological
constant at all. It appears only if
one is interested in interpreting results in terms of
the de Sitter spacetime or in the Poincare limit.
Since all the
operators $M^{ab}$ are dimensionless in units $\hbar/2=c=1$,
the de Sitter invariant quantum theories can be formulated
only in terms of dimensionless variables. As noted in Sect. \ref{S1},
this is a necessary requirement for a theory, which is supposed to
have a physical generalization to the case of Galois fields.
At the same time, since Poincare invariant theories do not have such
generalizations, one might expect that quantities which are dimensionful in
units $\hbar/2=c=1$ are not fundamental. In particular one might expect that 
the gravitational and cosmological constants are not fundamental. 
This is in the spirit of Mirmovich's hypothesis \cite{Mirmovich} that only
quantities having the dimension of the angular momentum can be fundamental. 

If one assumes that spacetime is fundamental then in the
spirit of General Relativity it is natural to think that
the empty space is flat, i.e. that the cosmological
constant is equal to zero. This was the subject of the
well-known dispute between Einstein and de Sitter
described in a wide literature. In QFT the cosmological constant
is given by a contribution of vacuum diagrams,
and the problem is to explain why it is so small. On the
other hand, if we assume that symmetry on quantum level in
our formulation is more fundamental then 
the cosmological constant problem does not arise at all. 
Instead we
have a problem of why nowadays Poincare symmetry is so
good approximate symmetry. This is rather a problem of cosmology but
not quantum physics. 

If a modular IR is considered in a linear space over
$F_{p^2}$ with $p=3\,\, (mod\,\, 4)$ then Eq. 
(\ref{8}) is also valid. However, as noted in Sect.
\ref{S1}, we consider modular IRs in linear spaces
over $F_{p^k}$ where $k$ is arbitrary. 
In this case it is
convenient to work with another set of ten operators.
Let $(a_j',a_j",h_j)$ $(j=1,2)$ be two independent sets
of operators satisfying the
commutation relations for the sp(2) algebra
\begin{equation}
[h_j,a_j']=-2a_j'\quad [h_j,a_j"]=2a_j"\quad [a_j',a_j"]=h_j
\label{9}
\end{equation}
The sets are independent in the sense that
for different $j$ they mutually commute with each other.
We denote additional four operators as $b', b",L_+,L_-$.
The operators
$L_3=h_1-h_2,L_+,L_-$ satisfy the commutation relations
of the su(2) algebra
\begin{equation}
[L_3,L_+]=2L_+\quad [L_3,L_-]=-2L_-\quad [L_+,L_-]=L_3
\label{10}
\end{equation}
while the other commutation relations are as follows
\begin{eqnarray}
&[a_1',b']=[a_2',b']=[a_1",b"]=[a_2",b"]=\nonumber\\
&[a_1',L_-]=[a_1",L_+]=[a_2',L_+]=[a_2",L_-]=0\nonumber\\
&[h_j,b']=-b'\quad [h_j,b"]=b"\quad
[h_1,L_{\pm}]=\pm L_{\pm},\nonumber\\
&[h_2,L_{\pm}]=\mp L_{\pm}\quad [b',b"]=h_1+h_2\nonumber\\
&[b',L_-]=2a_1'\quad [b',L_+]=2a_2'\quad [b",L_-]=-2a_2"\nonumber\\
&[b",L_+]=-2a_1",\quad [a_1',b"]=[b',a_2"]=L_-\nonumber\\
&[a_2',b"]=[b',a_1"]=L_+,\quad [a_1',L_+]=[a_2',L_-]=b'\nonumber\\
&[a_2",L_+]=[a_1",L_-]=-b"
\label{11}
\end{eqnarray}
At first glance these relations might seem rather
chaotic but in fact they are very natural in the Weyl basis
of the so(2,3) algebra.

In spaces over $F_{p^2}$ with $p=3\,\, (mod\,\, 4)$ the 
relation between the above sets of ten operators is
\begin{eqnarray}
&M_{10}=i(a_1"-a_1'-a_2"+a_2')\quad M_{15}=a_2"+a_2'-a_1"-a_1'\nonumber\\
&M_{20}=a_1"+a_2"+a_1'+a_2'\quad M_{25}=i(a_1"+a_2"-a_1'-a_2')\nonumber\\
&M_{12}=L_3\quad M_{23}=L_++L_-\quad M_{31}=-i(L_+-L_-)\nonumber\\
&M_{05}=h_1+h_2\quad M_{35}=b'+b"\quad M_{30}=-i(b"-b')
\label{12}
\end{eqnarray}
and therefore the sets are equivalent. However, the relations
(\ref{9}-\ref{11}) are more general since they can be used
when the representation space is a space over $F_{p^k}$ with
an arbitrary $k$.

We use the basis in which the operators
$(h_j,K_j)$ $(j=1,2)$ are diagonal. Here $K_j$ is the
Casimir operator (\ref{3}) for algebra $(a_j',a_j",h_j)$.
For constructing IRs we need operators relating different
representations of the sp(2)$\times$sp(2) algebra.
By analogy with Refs. \cite{Evans,Braden}, one of the
possible choices is as follows
\begin{eqnarray}
&A^{++}=b"(h_1-1)(h_2-1)-a_1"L_-(h_2-1)-a_2"L_+(h_1-1)
+\nonumber\\
&a_1"a_2"b' \quad A^{+-}=L_+(h_1-1)-a_1"b'\nonumber\\
&A^{-+}=L_-(h_2-1)-a_2"b'\quad A^{--}=b'
\label{13}
\end{eqnarray}
We consider the action of these operators only on the
space of 'minimal'
sp(2)$\times$sp(2) vectors, i.e. such vectors $x$ that
$a_j'x=0$ for $j=1,2$, and $x$ is the eigenvector of the
operators $h_j$. If $x$ is a minimal vector such that
$h_jx=\alpha_jx$ then $A^{++}x$ is the minimal
eigenvector of the
operators $h_j$ with the eigenvalues $\alpha_j+1$, $A^{+-}x$ -
with the eigenvalues $(\alpha_1+1,\alpha_2-1)$,
$A^{-+}x$ - with the eigenvalues $(\alpha_1-1,\alpha_2+1)$,
and $A^{--}x$ - with the eigenvalues $\alpha_j-1$.

By analogy with Refs. \cite{Evans,Braden}, we require
the existence of the vector $e_0$ satisfying the conditions
\begin{eqnarray}
&a_j'e_0=b'e_0=L_+e_0=0\quad h_je_0=q_je_0\quad (j=1,2)
\label{15}
\end{eqnarray}
where $q_j\in F_p$, $|f(q_j)|\ll p$, $f(q_j)>0$ for
$j=1,2$ and $f(q_1-q_2)\geq 0$. 
It is well known (see e.g. Ref. \cite{hep}) that 
$M^{05}=h_1+h_2$ is the AdS analog of the energy operator.
As follows from
Eqs. (\ref{9}) and (\ref{11}), the operators
$(a_1',a_2',b')$ reduce the AdS energy by two units.
Therefore $e_0$ is an analog the state with the minimum energy
which can be called the rest state, and the spin in our
units is equal to the maximum value
of the operator $L_3=h_1-h_2$ in that state. For these
reasons we use $s$ to denote $q_1-q_2$
and $m$ to denote $q_1+q_2$. 
In the standard classification
\cite{Evans}, the massive case is characterized by
the condition $q_2>1$ and massless one --- by
the condition $q_2=1$. There also exist two
exceptional IRs discovered by Dirac \cite{DiracS}
(Dirac singletons). As shown in Ref. \cite{lev2},
the modular analog of Dirac singletons is simple
and the massless case has been discussed in
detail in Ref. \cite{tmf}. For these reasons in
the present paper we consider only the massive case.

As follows from the above remarks, the elements
\begin{equation}
e_{nk}=(A^{++})^n(A^{-+})^ke_0
\label{16}
\end{equation}
represent the minimal sp(2)$\times$sp(2) vectors with the
eigenvalues of the operators $h_1$ and $h_2$ equal to
$Q_1(n,k)=q_1+n-k$ and $Q_2(n,k)=q_2+n+k$, respectively.
It can be shown by a direct calculation that
\begin{equation}
A^{--}A^{++}e_{nk}=(n+1)(m+n-2)(q_1+n)(q_2+n-1)e_{nk}
\label{17}
\end{equation}
\begin{equation}
A^{+-}A^{-+}e_{nk}=(k+1)(s-k)(q_1-k-2)(q_2+k-1)e_{nk}
\label{18}
\end{equation}

As follows from these expressions, in the massive
case $k$ can assume only 
the values $0,1,...s$ and in the standard theory
$n=0,1,...\infty$. 
However, in the modular case
$n=0,1,...n_{max}$ where $n_{max}$ is the first number
for which the r.h.s. of Eqs. (\ref{17}) becomes
zero in $F_p$. Therefore $n_{max}=p+2-m$.

The full basis of the representation space can 
be chosen in the form
\begin{equation}
e(n_1n_2nk)=(a_1")^{n_1}(a_2")^{n_2}e_{nk}
\label{19}
\end{equation}
In the standard theory $n_1$ and $n_2$ can be any 
natural numbers. However, as follows from the
results of the preceding section, Eq. (\ref{9}) and
the properties of the $A$ operators,
\begin{eqnarray}
&n_1=0,1,...N_1(n,k)\quad n_2=0,1,...N_2(n,k)\nonumber\\
&N_1(n,k)=p-q_1-n+k\quad N_2(n,k)=p-q_2-n-k
\label{20}
\end{eqnarray}
As a consequence, the representation is finite 
dimensional in agreement with the Zassenhaus 
theorem \cite{Zass} (moreover, it is
finite since any Galois field is finite).

Let us assume additionally that the representation
space is supplied by a scalar product (see Sect. \ref{S1}).
The element $e_0$ can always be chosen such
that $(e_0,e_0)=1$. Suppose that the representation 
operators satisfy the Hermiticity conditions
$L_+^*=L_-$, $a_j^{'*}=a_j"$, $b^{'*}=b"$ and $h_j^*=h_j$.
Then, as follows from Eq. (\ref{12}), in a special case
when the representation space is a space  
over $F_{p^2}$ with $p=3\,\, (mod\,\, 4)$, the operators 
$M^{ab}$ are Hermitian as it should be. By using Eqs. 
(\ref{9}-\ref{18}), one can show by a direct calculation
that the elements $e(n_1n_2nk)$ are mutually orthogonal
while the quantity
\begin{equation}
Norm(n_1n_2nk)=(e(n_1n_2nk),e(n_1n_2nk))
\label{23}
\end{equation}
can be represented as
\begin{equation}
Norm(n_1n_2nk)=F(n_1n_2nk)G(nk)
\label{24}
\end{equation}
where
\begin{eqnarray}
&F(n_1n_2nk)= n_1!(Q_1(n,k)+n_1-1)!n_2!(Q_2(n,k)+n_2-1)!\nonumber\\
&G(nk)=\{(q_2+k-2)!n!(m+n-3)!(q_1+n-1)!\nonumber\\
&(q_2+n-2)!k!s!\}\{(q_1-k-2)![(q_2-2)!]^3(q_1-1)!\nonumber\\
&(m-3)!(s-k)![Q_1(n,k)-1][Q_2(n,k)-1]\}^{-1}
\label{25}
\end{eqnarray}

In standard Poincare and AdS theories there also exist IRs with
negative energies. They can be constructed by analogy with 
positive energy IRs.
Instead of Eq. (\ref{15}) one can require the existence of the
vector $e_0'$ such that
\begin{eqnarray}
&a_j"e_0'=b"e_0'=L_-e_0'=0\quad h_je_0'=-q_je_0'\nonumber\\
&(e_0',e_0')\neq 0\quad (j=1,2)
\label{26}
\end{eqnarray}
where the quantities $q_1,q_2$ are the same as for positive
energy IRs. It is obvious that positive and negative energy
IRs are fully independent since the spectrum of the operator
$M^{05}$ for such IRs is positive and negative, respectively.
However, {\it the modular analog 
of a positive energy IR characterized by $q_1,q_2$ in
Eq. (\ref{15}), and the modular
analog of a negative energy IR characterized by the same
values of $q_1,q_2$ in Eq. (\ref{26}) represent the same
modular IR.} This is the crucial difference between the
standard quantum theory and GFQT, and a proof is given
below.

\begin{sloppypar}
Let $e_0$ be a vector satisfying Eq. (\ref{15}). Denote
$N_1=p-q_1$ and $N_2=p-q_2$. Our goal is to prove
that the vector $x=(a_1")^{N_1}(a_2")^{N_2}e_0$ satisfies
the conditions
(\ref{26}), i.e. $x$ can be identified with $e_0'$.
\end{sloppypar}

As follows from the definition of $N_1,N_2$,
the vector $x$ is the eigenvector of the operators $h_1$
and $h_2$ with the eigenvalues $-q_1$ and $-q_2$,
respectively, and, in
addition, it satisfies the conditions $a_1"x=a_2"x=0$.
Let us prove that $b"x=0$. Since $b"$ commutes with the
$a_j"$, we can write $b"x$ in the form
\begin{equation}
b"x = (a_1")^{N_1}(a_2")^{N_2}b"e_0
\label{27}
\end{equation}
As follows from Eqs. (\ref{11}) and (\ref{15}),
$a_2'b"e_0=L_+e_0=0$ and $b"e_0$ is the eigenvector
of the operator $h_2$ with the eigenvalue $q_2+1$.
Therefore, $b"e_0$ is the minimal vector of the sp(2)
IR which has the dimension $p-q_2=N_2$.
Therefore $(a_2")^{N_2}b"e_0=0$ and $b"x=0$.

The next stage of the proof is to show that $L_-x=0$.
As follows from Eq. (\ref{11}) and the definition of
$x$,
\begin{equation}
L_-x = (a_1")^{N_1}(a_2")^{N_2}L_-e_0-
N_1(a_1")^{N_1-1}(a_2")^{N_2}b"e_0
\label{28}
\end{equation}
We have already shown that $(a_2")^{N_2}b"e_0=0$,
and therefore it is sufficient to prove that the first term
in the r.h.s. of Eq. (\ref{28}) is equal to zero. As follows
from Eqs. (\ref{11}) and (\ref{15}), $a_2'L_-e_0=b'e_0=0$,
and $L_-e_0$ is the eigenvector of the operator $h_2$ with the
eigenvalue $q_2+1$. Therefore $(a_2")^{N_2}L_-e_0=0$ 
and the proof is completed.

Let us assume for a moment that the eigenvalues of 
the operators $h_1$ and $h_2$ should be treated 
not as elements 
of $F_p$ but as integers. Then, as follows from
the consideration in the preceding section, one 
modular IR of the so(2,3) algebra corresponds 
to a standard
positive energy IR in the region where the energy is
positive and much less than $p$. At the same time, it
corresponds to an IR with the negative energy in the
region where the AdS energy is close to $4p$ but less
than $4p$. This observation will be used in the
discussion of the vacuum condition in Sect. \ref{S5}.

The matrix elements of the operator $A$ are defined as
\begin{equation}
Ae(n_1n_2nk)=\sum_{n_1'n_2'n'k'}
A(n_1'n_2'n'k';n_1n_2nk)e(n_1'n_2'n'k')
\label{29}
\end{equation}
where the sum is taken over all possible values of
$(n_1'n_2'n'k')$. One can explicitly calculate 
matrix elements for all the representation operators 
and the results are as follows.
\begin{eqnarray}
&h_1e(n_1n_2nk)=[Q_1(n,k)+2n_1]e(n_1n_2nk)\nonumber\\
&h_2e(n_1n_2nk)=[Q_2(n,k)+2n_2]e(n_1n_2nk)
\label{30}
\end{eqnarray}
\begin{eqnarray}
&a_1'e(n_1n_2nk)=n_1[Q_1(n,k)+n_1-1]e(n_1-1,n_2nk)\nonumber\\
&a_1"e(n_1n_2nk)=e(n_1+1,n_2nk)\nonumber\\
&a_2'e(n_1n_2nk)=n_2[Q_2(n,k)+n_2-1]e(n_1,n_2-1,nk)\nonumber\\
&a_2"e(n_1n_2nk)=e(n_1,n_2+1,nk)
\label{31}
\end{eqnarray}
\begin{eqnarray}
&b"e(n_1n_2nk)=\{[Q_1(n,k)-1][Q_2(n,k)-1]\}^{-1}\nonumber\\
&[k(s+1-k)(q_1-k-1)(q_2+k-2)e(n_1,n_2+1,n,k-1)+\nonumber\\
&n(m+n-3)(q_1+n-1)(q_2+n-2)e(n_1+1,n_2+1,n-1,k)+\nonumber\\
&e(n_1,n_2,n+1,k)+e(n_1+1,n_2,n,k+1)]
\label{32}
\end{eqnarray}
\begin{eqnarray}
&b'e(n_1n_2nk)=\{[Q_1(n,k)-1][Q_2(n,k)-1]\}^{-1}
[n(m+n-3)\nonumber\\
&(q_1+n-1)(q_2+n-2)(q_1+n-k+n_1-1)(q_2+n+k+n_2-1)\nonumber\\
&e(n_1n_2,n-1,k)+n_2(q_1+n-k+n_1-1)e(n_1,n_2-1,n,k+1)+\nonumber\\
&n_1(q_2+n+k+n_2-1)k(s+1-k)(q_1-k-1)(q_2+k-2)\nonumber\\
&e(n_1-1,n_2,n,k-1)+n_1n_2e(n_1-1,n_2-1,n+1,k)]
\label{33}
\end{eqnarray}
\begin{eqnarray}
&L_+e(n_1n_2nk)=\{[Q_1(n,k)-1][Q_2(n,k)-1]\}^{-1}
\{(q_2+n+k+\nonumber\\
&n_2-1)[k(s+1-k)(q_1-k-1)(q_2+k-2)e(n_1n_2n,k-1)+\nonumber\\
&n(m+n-3)(q_1+n-1)(q_2+n-2)e(n_1+1,n_2,n-1,k)]+\nonumber\\
&n_2[e(n_1,n_2-1,n+1,k)+e(n_1+1,n_2-1,n,k+1)]\}
\label{34}
\end{eqnarray}
\begin{eqnarray}
&L_-e(n_1n_2nk)=\{[Q_1(n,k)-1][Q_2(n,k)-1]\}^{-1}
\{n_1[k(s+1-k)\nonumber\\
&(q_1-k-1)(q_2+k-2)e(n_1-1,n_2n,k-1)+e(n_1-1,n_2,\nonumber\\
&n+1,k)]+(q_1+n-k+n_1-1)[e(n_1n_2n,k+1)+n(m+n-3)\nonumber\\
&(q_1+n-1)(q_2+n-2)e(n_1,n_2+1,n-1,k)]\}
\label{35}
\end{eqnarray}
We will always use a convention that $e(n_1n_2nk)$ is a null
vector if some of the numbers $(n_1n_2nk)$ are not in the 
range described above.  

The important difference between the standard and modular 
IRs is that in the latter the trace of each representation
operator is equal to zero while in the former this is 
obviouly not the
case (for example, the energy operator is positive definite
for IRs defined by Eq. (\ref{15}) and negative definite
for IRs defined by Eq. (\ref{26})).
For the operators $(a_j',a_j",L_{\pm},b',b")$ the validity
of this statement is clear immediately: since they
necessarily change one of the quantum numbers $(n_1n_2nk)$,
they do not contain
nonzero diagonal elements at all. The proof for the
diagonal operators $h_1$ and $h_2$ is as follows. For each
IR of the sp(2) algebra with the 'minimal weight' $q_0$ and
the dimension $N+1$, the eigenvalues of the operator $h$ are
$(q_0,q_0+2,...q_0+2N)$. The sum of these eigenvalues is
equal to zero in $F_p$ since $q_0+N=0$ in $F_p$ (see the
preceding section). Therefore we conclude that for any
representation operator $A$ 
\begin{equation}
\sum_{n_1n_2nk} A(n_1n_2nk,n_1n_2nk)=0
\label{36}
\end{equation}
This property is very important for investigating a new
symmetry between particles and antiparticles in the GFQT
which is discussed in the subsequent section.

\section{Quantization and AB symmetry}
\label{S4}

Let us first recall how the Fock space is
defined in the standard theory. Let $a(n_1n_2nk)$ be the
operator of particle annihilation in the state described
by the vector $e(n_1n_2nk)$. Then the adjoint operator
$a(n_1n_2nk)^*$ has the meaning of particle creation in
that state. Since we do not normalize the states
$e(n_1n_2nk)$ to one, we require that the operators
$a(n_1n_2nk)$ and $a(n_1n_2nk)^*$ should satisfy
either the anticommutation relations
\begin{eqnarray}
&\{a(n_1n_2nk),a(n_1'n_2'n'k')^*\}=\nonumber\\
&Norm(n_1n_2nk)
\delta_{n_1n_1'}\delta_{n_2n_2'}\delta_{nn'}\delta_{kk'}
\label{37}
\end{eqnarray}
or the commutation relations
\begin{eqnarray}
&[a(n_1n_2nk),a(n_1'n_2'n'k')^*]=\nonumber\\
&Norm(n_1n_2nk)
\delta_{n_1n_1'}\delta_{n_2n_2'}\delta_{nn'}\delta_{kk'}
\label{38}
\end{eqnarray}

In the standard theory the representation describing
a particle and its antiparticle are fully independent
and therefore quantization of antiparticles should
be described by other operators.
If $b(n_1n_2nk)$ and $b(n_1n_2nk)^*$ are
operators of the antiparticle annihilation and creation
in the state $e(n_1n_2nk)$ then by analogy with Eqs.
(\ref{37}) and (\ref{38})
\begin{eqnarray}
&\{b(n_1n_2nk),b(n_1'n_2'n'k')^*\}=\nonumber\\
&Norm(n_1n_2nk)
\delta_{n_1n_1'}\delta_{n_2n_2'}\delta_{nn'}\delta_{kk'}
\label{39}
\end{eqnarray}
\begin{eqnarray}
&[b(n_1n_2nk),b(n_1'n_2'n'k')^*]=\nonumber\\
&Norm(n_1n_2nk)
\delta_{n_1n_1'}\delta_{n_2n_2'}\delta_{nn'}\delta_{kk'}
\label{40}
\end{eqnarray}
for anticommutation or commutation relations,
respectively.
In this case it is assumed that in the case of anticommutation
relations all the operators $(a,a^*)$ anticommute with
all the operators $(b,b^*)$ while in the case of
commutation relations they commute with each other. It is
also assumed that the Fock space contains the vacuum vector
$\Phi_0$ such that
\begin{equation}
a(n_1n_2nk)\Phi_0=b(n_1n_2nk)\Phi_0=0\quad
\forall\,\, n_1,n_2,n,k
\label{41}
\end{equation}

The Fock space in the standard theory can now be defined as 
a linear combination of all elements obtained by the action 
of the operators
$(a^*,b^*)$ on the vacuum vector, and the problem of
second quantization of representation
operators can
be formulated as follows. Let $(A_1,A_2....A_n)$
be representation
operators describing IR of the AdS algebra. One should
replace them by operators acting in the Fock space
such that the commutation relations between their
images in the Fock space are the same as for original
operators (in other words, we should have a homomorphism
of Lie algebras of operators acting in the space of IR
and in the Fock space). We can also require that our
map should be compatible with the Hermitian
conjugation in both spaces. It is easy to verify that
a possible solution satisfying all the requirements is
as follows. Taking into account the fact that the
matrix elements satisfy the proper commutation relations,
the operators $A_i$ in the quantized form
\begin{eqnarray}
&A_i=\sum A_i(n_1'n_2'n'k',n_1n_2nk)
[a(n_1'n_2'n'k')^*a(n_1n_2nk)+\nonumber\\
&b(n_1'n_2'n'k')^*b(n_1n_2nk)]/Norm(n_1n_2nk)
\label{42}
\end{eqnarray}
satisfy the commutation relations (\ref{9}-\ref{11}). 
We will not use
special notations for operators in the Fock space since
in each case it will be clear whether the operator in
question acts in the space of IR or in the Fock space.

\begin{sloppypar}
A well known problem in the standard theory is that
the quantization procedure does not define the order
of the annihilation and creation operators uniquely.
For example, another possible solution is
\begin{eqnarray}
&A_i=\mp \sum A_i(n_1'n_2'n'k',n_1n_2nk)
[a(n_1n_2nk)a(n_1'n_2'n'k')^*+\nonumber\\
&b(n_1n_2nk)b(n_1'n_2'n'k')^*]/Norm(n_1n_2nk)
\label{43}
\end{eqnarray}
for anticommutation and commutation
relations, respectively. The solutions (\ref{42}) and
(\ref{43}) are different since the energy operators
$M^{05}$ in these expressions differ by an infinite
constant. In the standard theory the solution 
(\ref{42}) is selected by imposing an additional
requirement that all operators should be written 
in the normal form where annihilation operators 
precede creation ones. Then the vacuum 
has zero energy
and Eq. (\ref{43}) should be rejected. Such a
requirement does not follow from the
theory. Ideally there should be a procedure which 
correctly defines the order of operators from first
principles. 
\end{sloppypar}

In the standard theory there also exist neutral
particles. In that case there is no need to have two
independent sets of operators $(a,a^*)$ and $(b,b^*)$,
and Eq. (\ref{42}) should be written without the
$(b,b^*)$ operators. The problem of neutral particles
in GFQT is discussed in Sect. \ref{S7}.

We now proceed to quantization in the modular case.
The results of the preceding section
show that one modular IR corresponds to two standard
IRs with the positive and negative energies,
respectively. This indicates to a possibility that
one modular IR describes a particle and its
antiparticle simultaneously. However, we don't
know yet what should be treated as a particle and
its antiparticle in the modular case. We have a
description of an object such that $(n_1n_2nk)$ is
the full set of its quantum numbers which take 
the values described in the preceding section. 

We now assume that
$a(n_1n_2nk)$ in GFQT is the operator describing
annihilation of the object with the quantum
numbers $(n_1n_2nk)$ regardless of whether the
numbers are physical or nonphysical. 
Analogously $a(n_1n_2nk)^*$
describes creation of the object with the quantum
numbers $(n_1n_2nk)$. If these operators anticommute
then they satisfy Eq. (\ref{37}) while if they
commute then they satisfy Eq. (\ref{38}). 
Then, by analogy with the standard case, the operators
\begin{eqnarray}
&A_i=\sum A_i(n_1'n_2'n'k',n_1n_2nk)\nonumber\\
&a(n_1'n_2'n'k')^*a(n_1n_2nk)/Norm(n_1n_2nk)
\label{44}
\end{eqnarray}
satisfy the commutation relations (\ref{9}-\ref{11}). 
In this expression
the sum is taken over all possible values of the
quantum numbers in the modular case.

In the modular case the solution can be taken
not only as in Eq. (\ref{44}) but also as
\begin{eqnarray}
&A_i=\mp\sum A_i(n_1'n_2'n'k',n_1n_2nk)\nonumber\\
&a(n_1n_2nk)a(n_1'n_2'n'k')^*/Norm(n_1n_2nk)
\label{45}
\end{eqnarray}
for the cases of anticommutators and commutators,
respectively.
However, as follows from Eqs. (\ref{36}-\ref{38}),
the solutions (\ref{44}) and (\ref{45}) are the 
same. Therefore in the
modular case there is no need to impose an
artificial requirement that all operators
should be written in the normal form.

The problem with the treatment of the $(a,a^*)$
operators is as follows.  When the values of
$(n_1n_2n)$ are much less than $p$, the modular IR
corresponds to the standard one and therefore 
the $(a,a^*)$ operator can
be treated as those describing the particle
annihilation and creation, respectively. However,
when the AdS energy is negative, the operators
$a(n_1n_2nk)$ and $a(n_1n_2nk)^*$ become unphysical
since they describe annihilation and creation,
respectively, in the unphysical region of negative
energies.

Let us recall that at any fixed
values of $n$ and $k$, the quantities $n_1$ and $n_2$
can take only the values described in Eq. (\ref{20})
and the eigenvalues of the operators $h_1$ and $h_2$
are given by $Q_1(n,k)+2n_1$ and $Q_2(n,k)+2n_2$,
respectively.
As follows from the results of the preceding section,
the first IR of the sp(2) algebra has the dimension
$N_1(n,k)+1$ and the second IR has the dimension
$N_2(n,k)+1$. If $n_1=N_1(n,k)$
then it follows from Eq. (\ref{20}) that the first
eigenvalue is equal to $-Q_1(n,k)$ in $F_p$, and if
$n_2=N_2(n,k)$ then
the second eigenvalue is equal to $-Q_2(n,k)$ in $F_p$.
We use ${\tilde n}_1$ to denote $N_1(n,k)-n_1$ and
${\tilde n}_2$ to denote
$N_2(n,k)-n_2$. Then it follows from Eq.
(\ref{20}) that
$e({\tilde n}_1{\tilde n}_2nk)$ is the
eigenvector
of the operator $h_1$ with the eigenvalue $-(Q_1(n,k)+2n_1)$
and the
eigenvector of the operator $h_2$ with the eigenvalue
$-(Q_2(n,k)+2n_2)$.

The standard theory
implicitly involves the idea that creation of the
antiparticle with the positive energy can be treated
as annihilation of the corresponding particle with the
negative energy and annihilation of the
antiparticle with the positive energy can be treated
as creation of the corresponding particle with the
negative energy. In GFQT we can implement this
idea explicitly. Namely, we can define the
operators $b(n_1n_2nk)$ and $b(n_1n_2nk)^*$ in such a
way that they will replace the $(a,a^*)$ operators if
the quantum numbers are unphysical. In addition,
if the values of $(n_1n_2n)$ are much less than $p$,
the operators $b(n_1n_2nk)$ and $b(n_1n_2nk)^*$
should be interpreted as physical operators
describing annihilation and creation of
antiparticles, respectively.

In GFQT the $(b,b^*)$ operators cannot be independent
of the $(a,a^*)$ operators since the latter are defined
for all possible quantum numbers. Therefore the $(b,b^*)$
operators should be expressed in terms of the $(a,a^*)$
ones. We can implement the above idea if the
operator $b(n_1n_2nk)$ is defined in such a way that
it is proportional to $a({\tilde n}_1,{\tilde n}_2,n,k)^*$
and hence $b(n_1n_2nk)^*$ is proportional to
$a({\tilde n}_1,{\tilde n}_2,n,k)$.

Since Eq. (\ref{25}) should now be considered in $F_p$,
it follows from the well known Wilson
theorem $(p-1)!=-1$ in $F_p$ (see e.g. \cite{VDW}) that
\begin{equation}
F(n_1n_2nk)F({\tilde n}_1{\tilde n}_2nk) = (-1)^s
\label{46}
\end{equation}
We now define the $b$-operators as
\begin{equation}
a(n_1n_2nk)^*=\eta(n_1n_2nk) b({\tilde n}_1{\tilde n}_2nk)/
F({\tilde n}_1{\tilde n}_2nk)
\label{47}
\end{equation}
where $\eta(n_1n_2nk)$ is some function.
As a consequence,
\begin{eqnarray}
&a(n_1n_2nk)=\bar{\eta}(n_1n_2nk) b({\tilde n}_1{\tilde n}_2nk)^*/
F({\tilde n}_1{\tilde n}_2nk)\nonumber\\
&b(n_1n_2nk)^*=a({\tilde n}_1{\tilde n}_2nk)
F(n_1n_2nk)/{\bar \eta}({\tilde n}_1{\tilde n}_2nk)\nonumber\\
&b(n_1n_2nk)=a({\tilde n}_1{\tilde n}_2nk)^*
F(n_1n_2nk)/\eta({\tilde n}_1{\tilde n}_2nk)
\label{48}
\end{eqnarray}

Eqs. (\ref{47}) and (\ref{48}) define a relation
between the sets $(a,a^*)$ and $(b,b^*)$. Although our
motivation was to replace the $(a,a^*)$ operators by the
$(b,b^*)$ ones only for the nonphysical values of the
quantum numbers, we can consider this definition for all
the values of $(n_1n_2nk)$.

We have not discussed yet what exact definition of the
physical and nonphysical quantum numbers should be. 
This problem
will be discussed in Sect. \ref{S5}. However, one might 
accept

{\it Physical-nonphysical states assumption:
Each set of quantum numbers $(n_1n_2nk)$
is either physical or unphysical. If it is physical then
the set $({\tilde n}_1{\tilde n}_2nk)$ is unphysical
and vice versa.}

With this assumption we can conclude from Eqs. (\ref{47})
and (\ref{48}) that if some operator $a$ is physical then
the corresponding operator $b^*$ is unphysical and vice
versa while if some operator $a^*$ is physical then
the corresponding operator $b$ is unphysical and vice
versa.

We have no ground to think that the set of the $(a,a^*)$
operators is more fundamental than the set of the $(b,b^*)$
operators and vice versa. Therefore the question arises
whether the $(b,b^*)$ operators satisfy the relations
(\ref{38}) or (\ref{39}) in the case of anticommutation
or commutation relations, respectively and whether the
operators $A_i$ (see Eq. (\ref{44})) have the same form
in terms of the $(a,a^*)$ and $(b,b^*)$ operators. In 
other words, if the $(a,a^*)$ operators in Eq. (\ref{44})
are expressed in terms of the $(b,b^*)$ ones then the
problem arises whether
\begin{eqnarray}
&A_i=\sum A_i(n_1'n_2'n'k',n_1n_2nk)\nonumber\\
&b(n_1'n_2'n'k')^*b(n_1n_2nk)/Norm(n_1n_2nk)
\label{49}
\end{eqnarray}
is valid. It is natural to accept the following

{\it Definition of the AB symmetry: If the $(b,b^*)$
operators satisfy Eq. (\ref{39}) in the case of
anticommutators or Eq. (\ref{40}) in the case of
commutators and all the representation operators 
(\ref{44}) in terms of the $(b,b^*)$ operators have 
the form (\ref{49}) then
it is said that the AB symmetry is satisfied.}

To prove the AB symmetry we will first investigate 
whether Eqs. (\ref{39})
and (\ref{40}) follow from Eqs. (\ref{37}) and
(\ref{38}), respectively. As follows from Eqs.
(\ref{46}-\ref{48}), Eq. (\ref{39}) follows from
Eq. (\ref{37}) if
\begin{equation}
\eta(n_1n_2nk) {\bar \eta}(n_1,n_2,nk)=(-1)^s
\label{50}
\end{equation}
while Eq. (\ref{40}) follows from Eq. (\ref{38}) if
\begin{equation}
\eta(n_1n_2nk) {\bar \eta}(n_1,n_2,nk)=(-1)^{s+1}
\label{51}
\end{equation}
We now represent $\eta(n_1n_2nk)$ in the form
\begin{equation}
\eta(n_1n_2nk)=\alpha f(n_1n_2nk)
\label{52}
\end{equation}
where $f(n_1n_2nk)$ should satisfy the condition
\begin{equation}
f(n_1n_2nk) {\bar f}(n_1,n_2,nk)=1
\label{53}
\end{equation}
Then $\alpha$ should be such that
\begin{equation}
\alpha {\bar \alpha}=\pm (-1)^s
\label{54}
\end{equation}
where the plus sign refers to anticommutators and the minus
sign to commutators, respectively.
If the normal spin-statistics connection is valid,
i.e. we have anticommutators for odd values of $s$ and
commutators for even ones then the r.h.s. of Eq.
(\ref{54}) equals -1 while in the opposite case it
equals 1. In Sect. \ref{S7}, Eq. (\ref{54}) is
discussed in detail and for now we assume that solutions
of this relation exist.

A direct calculation using the explicit expressions 
(\ref{30}-\ref{35}) for the matrix elements shows
that if $\eta(n_1n_2nk)$ is given by
Eq. (\ref{52}) and  
\begin{equation}
f(n_1n_2nk)=(-1)^{n_1+n_2+n}
\label{55}
\end{equation}
then the AB symmetry is valid
regardless of whether the normal spin-statistics
connection is valid or not (the details of calculations 
can be found in Ref. \cite{hep}). 

\section{Physical and nonphysical states}
\label{S5}

\begin{sloppypar}
The operator $a(n_1n_2nk)$ can be the physical
annihilation operator only if it annihilates the vacuum
vector $\Phi_0$. Then if the operators $a(n_1n_2nk)$ and
$a(n_1n_2nk)^*$ satisfy the relations (\ref{37}) or
(\ref{38}), the vector $a(n_1n_2nk)^* \Phi_0$ has the
meaning of the one-particle state. The same can be said
about the operators $b(n_1n_2nk)$ and $b(n_1n_2nk)^*$.
For these reasons in the standard theory it is required
that the vacuum vector should satisfy the conditions
(\ref{41}). Then the elements
\begin{equation}
\Phi_+(n_1n_2nk)=a(n_1n_2nk)^*\Phi_0\quad
\Phi_-(n_1n_2nk)=b(n_1n_2nk)^*\Phi_0
\label{56}
\end{equation}
have the meaning of one-particle states for particles
and antiparticles, respectively.
\end{sloppypar}

However, if one requires the condition (\ref{41})
in GFQT, then it is obvious from Eqs. (\ref{47})
and Eq. (\ref{48}) that the elements defined by
Eq. (\ref{56}) are null vectors. Note that in the
standard approach the AdS energy is always greater
than $m$ while in the GFQT the AdS energy is not
positive definite. We can therefore try to modify
Eq. (\ref{41}) as follows. Suppose that
'Physical-nonphysical states assumption' (see Sect.
\ref{S4}) can be substantiated. Then we can break
the set of elements $(n_1n_2nk)$ into two equal
nonintersecting parts, $S_+$ and $S_-$, such that if
$(n_1n_2nk)\in S_+$ then 
$({\tilde n}_1{\tilde n}_2nk)\in S_-$
and vice versa. Then, instead of the condition
(\ref{41}) we require
\begin{equation}
a(n_1n_2nk)\Phi_0=b(n_1n_2nk)\Phi_0=0\quad
\forall\,\, (n_1,n_2,n,k)\in S_+
\label{57}
\end{equation}
In that case the elements defined by Eq. (\ref{56})
will indeed have the meaning of one-particle states
for $(n_1n_2nk)\in S_+$.

It is clear that if we
wish to work with the full set of elements
$(n_1n_2nk)$ then, as follows from Eqs. (\ref{47})
and (\ref{48}), the operators $(b,b^*)$ are redundant
and we can work only with the operators $(a,a^*)$.
However, if one works with the both sets, $(a,a^*)$
and $(b,b^*)$ then such operators can be independent of
each other only for a half of the elements $(n_1n_2nk)$.

\begin{sloppypar}
Regardless of how the sets $S_+$ and $S_-$ are defined,
the 'Physical-nonphysical states assumption' cannot be
consistent if there exist quantum numbers $(n_1n_2nk)$
such that $n_1={\tilde n}_1$ and $n_2={\tilde n}_2$.
Indeed, in that case the sets $(n_1n_2nk)$ and
$({\tilde n}_1{\tilde n}_2nk)$ are the same what contradicts
the assumption that each set $(n_1n_2nk)$ belongs either
to $S_+$ or $S_-$.
\end{sloppypar}

Since the replacements $n_1\rightarrow {\tilde n}_1$ and
$n_2\rightarrow {\tilde n}_2$ change the signs of the
eigenvalues of the $h_1$ and $h_2$ operators (see Sect.
\ref{S4}), the condition that that $n_1={\tilde n}_1$
and $n_2={\tilde n}_2$ should be valid simultaneously
implies that the eigenvalues of the operators $h_1$ and
$h_2$ should be equal to zero simultaneously. Recall that
(see Sect. \ref{S2}) if one considers IR of the sp(2)
algebra and treats the eigenvalues of the diagonal operator
$h$ not as elements of $F_p$ but as integers, then they
take the values of $q_0,q_0+2,...2p-q_0-2,2p-q_0$. Therefore
the eigenvalue is equal to zero in $F_p$ only if it
is equal to $p$ when considered as an integer. Since the
AdS energy is $E=h_1+h_2$, the above situation can take 
place only if the energy
considered as an integer is equal to 2p. It now follows from
Eq. (\ref{12}) that the energy can be equal to $2p$ only
if $m$ is even. Since $m=q_1+q_2$ and $s=q_1-q_2$, we conclude
that $m$ can be even if and only if $s$ is even. In that case
we will necessarily have quantum numbers $(n_1n_2nk)$ such
that the sets $(n_1n_2nk)$ and $({\tilde n}_1{\tilde n}_2nk)$
are the same and therefore the 'Physical-nonphysical states
assumption' is not valid. On the other hand, if $s$ is odd
(i.e. half-integer in the usual units) then there are no
quantum numbers $(n_1n_2nk)$ such that the sets $(n_1n_2nk)$
and $({\tilde n}_1{\tilde n}_2nk)$ are the same.

Our conclusion is as follows: {\it If the separation of states
should be valid for any quantum numbers then the spin $s$ should
be necessarily odd.} In other words, if the notion of particles
and antiparticles is absolute then elementary particles
can have only a half-integer spin in the usual units.

In view of the above 
observations it seems natural to implement the 
'Physical-nonphysical states assumption'
as follows.
{\it If the quantum numbers $(n_1n_2nk)$ are such that
$m+2(n_1+n_2+n) < 2p$ then the corresponding state is
physical and belongs to $S_+$, otherwise the state is
unphysical and belongs to $S_-$.}

\section{Dirac vacuum energy problem}
\label{S6}

The Dirac vacuum energy problem is discussed in practically
every textbook on QFT. In its simplified form it can be
described as follows. Suppose that the
energy spectrum is discrete and $n$ is the quantum number
enumerating the states. Let $E(n)$ be the energy in the state
$n$. Consider the electron-positron field.
As a result of quantization one gets for the energy operator
\begin{equation}
E = \sum_n E(n)[a(n)^*a(n)-b(n)b(n)^*]
\label{58}
\end{equation}
where $a(n)$ is the operator of electron annihilation in the
state $n$, $a(n)^*$ is the operator of electron creation in
the state $n$, $b(n)$ is the operator of positron 
annihilation in the
state $n$ and $b(n)^*$ is the operator of positron creation in
the state $n$. It follows from this expression that only
anticommutation relations are possible since otherwise
the energy of positrons will be negative. However, if
anticommutation relations are assumed, it follows from
Eq. (\ref{58}) that
\begin{equation}
E = \{\sum_n E(n)[a(n)^*a(n)+b(n)^*b(n)]\}+E_0
\label{59}
\end{equation}
where $E_0$ is some infinite negative constant. Its presence
was a motivation for developing Dirac's hole
theory. In the modern approach it is usually required that the
vacuum energy should
be zero. This can be obtained by requiring that all
operators should be written in the normal form. 
This requirement does not follow from the theory, it is imposed with
the only purpose to obtain the correct result for the vacuum energy
and other observables. Also, as noted in Sect. \ref{S4}, this requirement 
is not quite consistent since the result of
quantization is Eq. (\ref{58}) where the positron operators
are not written in the normal form.

Consider now the AdS energy operator $M^{05}=h_1+h_2$ in
GFQT. As follows from Eqs. (\ref{30}) and (\ref{45})
\begin{eqnarray}
&M^{05}=\sum [m+2(n_1+n_2+n)]a(n_1n_2nk)^*\nonumber\\
&a(n_1n_2nk)/Norm(n_1n_2nk)
\label{60}
\end{eqnarray}
where the sum is taken over all possible quantum numbers
$(n_1n_2nk)$. We now wish to
replace only the nonphysical $(a,a^*)$
operators by the physical $(b,b^*)$ ones and represent
$M^{05}$ in terms of physical operators only.
As follows from Eqs. (\ref{46}-\ref{48}) and 
(\ref{52}-\ref{54})
\begin{eqnarray}
&M^{05}=\{\sum_{S_+} [m+2(n_1+n_2+n)]
[a(n_1n_2nk)^*a(n_1n_2nk)+\nonumber\\
&b(n_1n_2nk)^*b(n_1n_2nk)]/Norm(n_1n_2nk)\}+E_{vac}
\label{61}
\end{eqnarray}
where
\begin{equation}
E_{vac}=\mp \sum_{S_+}  [m+2(n_1+n_2+n)]
\label{62}
\end{equation}
in the cases when the $(b,b^*)$
operators anticommute and commute, respectively.
The quantity $E_{vac}$ has the meaning of the vacuum 
energy in GFQT and the goal of this section is to 
explicitly calculate this quantity. We assume for
definiteness that the operators anticommute.

Consider first the sum in Eq. (\ref{62}) when the values of
$n$ and $k$ are fixed. It is convenient to distinguish the
cases $s > 2k$ and $s<2k$ (recall that $s$ is assumed to be
odd). If $s > 2k$ then, as follows from Eq. (\ref{20}),
the maximum value of $n_1$ is such that $m+2(n+n_1)$ is always
less than $2p$. For this reason all the values of $n_1$
contribute to the sum, which can be written as
\begin{eqnarray}
&S_1(n,k) =-\sum_{n_1=0}^{p-q_1-n+k}[(m+2n+2n_1)+\nonumber\\
&(m+2n+2n_1+2)+...+(2p-1)]
\label{63}
\end{eqnarray}
A simple calculation shows that the result 
can be represented as
\begin{equation}
S_1(n,k)=\sum_{n_1=1}^{p-1}n_1^2-\sum_{n_1=1}^{n+(m-3)/2}n_1^2-
\sum_{n_1=1}^{(s-1)/2-k}n_1^2
\label{65}
\end{equation}
where the last sum should be taken into account only
if $(s-1)/2-k\geq 1$.

The first sum in this expression equals $(p-1)p(2p-1)/6$
and, since we assume that $p\neq 2$ and $p\neq 3$, this
quantity is zero in $F_p$. As a result, $S_1(n,k)$ is
represented as a sum of two terms such that the first one
depends only on $n$ and the second --- only on $k$. Note
also that the second term is absent if $s=1$, i.e. for
particles with the spin 1/2 in the usual units.

Analogously, if $s < 2k$ the result is
\begin{equation}
S_2(n,k)=-\sum_{n_2=1}^{n+(m-3)/2}n_2^2-\sum_{n_2=1}^{k-(s+1)/2}n_2^2
\label{67}
\end{equation}
where the second term should be taken into account only
if $k-(s+1)/2\geq 1$.

We now should calculate the sum
\begin{equation}
S(n)=\sum_{k=0}^{(s-1)/2}S_1(n,k) +\sum_{k=(s+1)/2}^s S_2(n,k)
\label{68}
\end{equation}
and the result is
\begin{eqnarray}
&S(n)=-(s+1)(n+\frac{m-1}{2})[2(n+\frac{m-1}{2})^2-\nonumber\\
&3(n+\frac{m-1}{2})+1]/6-(s-1)(s+1)^2(s+3)/96
\label{69}
\end{eqnarray}
Since the value of $n$ is in the range $[0,n_{max}]$,
the final result is
\begin{equation}
E_{vac}=\sum_{n=0}^{n_{max}}S(n)=(m-3)(s-1)(s+1)^2(s+3)/96
\label{70}
\end{equation}
since in the massive case $n_{max}=p+2-m$.

Our final conclusion in this section is that
{\it if $s$ is odd and the separation of states into physical
and nonphysical ones is accomplished as in Sect. \ref{S5} then
$E_{vac}=0$ only if $s=1$ (i.e. $s=1/2$
in the usual units)}.

\section{Neutral particles and spin-statistics theorem}
\label{S7}

The nonexistence of neutral elementary particles in
GFQT is one of the most striking differences
between GFQT and the standard theory. 
One could give the following definition of neutral
particle:
\begin{itemize}
\item i) it is a particle coinciding with its
antiparticle
\item ii) it is a particle which does not coincide
with its antiparticle but they have the same properties
\end{itemize}
In the standard theory only i) is meaningful since
neutral particles are described by real (not complex)
fields and this condition is required by Hermiticity.
One might think that the definition ii) is only academic
since if a particle and its antiparticle have the same
properties then they are indistinguishable and can be
treated as the same. However, the cases i) and ii)
are essentially different from the operator point of
view. In the case i) only the $(a,a^*)$ operators
are sufficient for  describing the operators (\ref{42}).
This is the reflection of the fact that the real
field has the number of degrees of freedom twice
as less as the complex field. On the other hand,
in the case ii) both $(a,a^*)$ and $(b,b^*)$
operators are required, i.e. in the standard theory
such a situation is described by a complex field.
Nevertheless, the case ii) seems to be rather odd:
it implies that there exists a quantum number
distinguishing a particle from its antiparticle
but this number is not manifested experimentally.
We now consider whether the conditions i) or ii) can
be implemented in GFQT.

Since each operator $a$
is proportional to some operator $b^*$ and vice versa
(see Eqs. (\ref{47}) and (\ref{48})), it is
clear that if the particles described by the
operators $(a,a^*)$ have some nonzero charge then
the particles described by the operators $(b,b^*)$
have the opposite charge and the number of operators
cannot be reduced. However, if all possible charges are
zero, one could try to implement i) by requiring that
each $b(n_1n_2nk)$ should be proportional to
$a(n_1n_2nk)$ and then $a(n_1n_2nk)$ will be
proportional to $a({\tilde n}_1,{\tilde n}_2,nk)^*$.
In this case the operators $(b,b^*)$ will not be
needed at all.

Suppose, for example, 
that the operators $(a,a^*)$ satisfy the commutation
relations (\ref{38}). In that case
the operators $a(n_1n_2nk)$ and $a(n_1'n_2'n'k')$
should commute if the sets $(n_1n_2nk)$ and
$(n_1'n_2'n'k')$ are not the same.
In particular, one should have $[a(n_1n_2nk),
a({\tilde n}_1{\tilde n}_2nk)]=0$ if either
$n_1\neq {\tilde n}_1$ or $n_2\neq {\tilde n}_2$.
On the other hand, if $a({\tilde n}_1{\tilde n}_2nk)$
is proportional to $a(n_1n_2nk)^*$, it follows from
Eq. (\ref{38}) that the commutator cannot be zero.
Analogously one can consider the case of anticommutators.

The fact that the number of operators cannot be
reduced is also clear from the observation that the
$(a,a^*)$ or $(b,b^*)$ operators describe an
irreducible representation in which the number of
states (by definition) cannot be reduced. Our 
conclusion
is that in GFQT the definition of neutral particle
according to i) is fully unacceptable.

Consider now whether it is possible to implement
the definition ii) in GFQT. Recall that we started from
the operators $(a,a^*)$ and defined the operators
$(b,b^*)$ by means of Eq. (\ref{47}). Then the
latter satisfy the same commutation or
anticommutation relations as the former and the AB
symmetry is valid. Does it mean that the particles described
by the operators $(b,b^*)$ are the same as the ones
described by the operators $(a,a^*)$? If one starts
from the operators $(b,b^*)$ then, by analogy with Eq.
(\ref{47}), the operators $(a,a^*)$ can be defined as
\begin{equation}
b(n_1n_2nk)^*=\eta'(n_1n_2nk) a({\tilde n}_1{\tilde n}_2nk)/
F({\tilde n}_1{\tilde n}_2nk)
\label{72}
\end{equation}
where $\eta'(n_1n_2nk)$ is some function. By analogy
with the consideration in Sect. \ref{S4} one
can show that
\begin{equation}
\eta'(n_1n_2nk)=\beta (-1)^{n_1+n_2+n}\quad
\beta {\bar \beta}=\mp 1
\label{73}
\end{equation}
where the minus sign refers to the normal
spin-statistics connection and the plus sign ---
to the broken one.

As follows from Eqs. (\ref{47}), (\ref{50}-\ref{53}),
(\ref{72}), (\ref{73})
and the definition of the quantities ${\tilde n}_1$
and ${\tilde n}_2$ in Sect. \ref{S4}, the relation
between the quantities $\alpha$ and $\beta$ is
$\alpha {\bar \beta}=1$.
Therefore, as follows from Eq. (\ref{73}), there
exist only two possibilities, $\beta = \mp \alpha$,
depending on whether the normal spin-statistics
connection is valid or not.
We conclude that the broken spin-statistics connection
implies that $\alpha{\bar \alpha}=\beta{\bar\beta}=1$
and $\beta=\alpha$ while the normal spin-statistics
connection implies that
$\alpha{\bar \alpha}=\beta{\bar\beta}=-1$
and $\beta=-\alpha$. 
In the first case solutions for 
$\alpha$ and $\beta$ obviously
exist (e.g. $\alpha = \beta = 1$) and the particle and 
its antiparticle can be
treated as neutral in the sense of the definition ii).
Since such a situation is clearly unphysical, one might
treat the spin-statistics theorem as a requirement
excluding neutral particles in the sense ii).

\begin{sloppypar}
We now consider another possible treatment of the
spin-statistics theorem, which seems to be much
more interesting. In the case of normal 
spin-statistics connection we have that
\begin{equation}
\alpha {\bar \alpha}=-1
\label{77}
\end{equation}
and the problem arises whether solutions of this 
relation exist. Such a relation is obviously 
impossible in the standard theory.
\end{sloppypar}

As noted in Sect. \ref{S1}, $-1$ is a quadratic 
residue in $F_p$ if $p=1\,\, (mod\,\, 4)$ and
a quadratic nonresidue in $F_p$ if 
$p=3\,\, (mod\,\, 4)$. For example, $-1$ is a
quadratic residue in $F_5$ since 
$2^2=-1\,\, (mod\,\, 5)$ but in $F_7$ there is
no element $a$ such that
$a^2=-1\,\, (mod\,\, 7)$.
We conclude that if 
$p=1\,\, (mod\,\, 4)$ then Eq. (\ref{77})
has solutions in $F_p$ and in that case 
the theory can be constructed without any
extension of $F_p$.

Consider now the case $p=3\,\, (mod\,\, 4)$.
Then Eq. (\ref{77}) has no solutions in $F_p$
and it is necessary to consider this equation
in an extention of $F_p$ (i.e. there
is no 'real' version of GFQT). The minimum extension
is obviously $F_{p^2}$ and therefore the
problem arises whether 
Eq. (\ref{77}) has solutions in $F_{p^2}$.

It is well known \cite{VDW} that any Galois
field without its zero element is a cyclic
multiplicative group. Let $r$ be a primitive
root, i.e. the element such that any nonzero
element of $F_{p^2}$ can be represented as
$r^k$ $(k=1,2,...,p^2-1)$. It is also well
known that the only nontrivial automorphism
of $F_{p^2}$ is 
$\alpha\rightarrow {\bar \alpha}=\alpha^p$.
Therefore if $\alpha =r^k$ then $\alpha{\bar \alpha}=
r^{(p+1)k}$. On the other hand, since $r^{(p^2-1)}=1$, 
$r^{(p^2-1)/2}=-1$. Therefore there exists at
least a solution with $k=(p-1)/2$.

Our conclusion is that {\it if $p=3\,\, (mod\,\, 4 )$
then the spin-statistics theorem implies that 
the field $F_p$ should necessarily
be  extended and the minimum possible extension
is $F_{p^2}$}. Therefore the spin-statistics
theorem can be treated as a requirement that
GFQT should be based on $F_{p^2}$ and the 
standard theory should be based on complex 
numbers.

Let us now discuss a different approach to
the AB symmetry. A desire to have 
operators which can be interpreted as those relating 
separately to particles and antiparticles is natural 
in view of our
experience in the standard approach. However, one
might think that in the spirit of GFQT there is no 
need to have separate operators for
particles and antiparticles since they are different states
of the same object. We can therefore 
reformulate the AB symmetry in terms of only $(a,a^*)$ operators
as follows. Instead of 
Eqs. (\ref{47}) and (\ref{48}), we consider a 
{\it transformation} defined as 
\begin{eqnarray}
&a(n_1n_2nk)^*\rightarrow \eta(n_1n_2nk) 
a({\tilde n}_1{\tilde n}_2nk)/
F({\tilde n}_1{\tilde n}_2nk)\nonumber\\ 
&a(n_1n_2nk)\rightarrow \bar{\eta}(n_1n_2nk) 
a({\tilde n}_1{\tilde n}_2nk)^*/
F({\tilde n}_1{\tilde n}_2nk)
\label{78}
\end{eqnarray}
Then the AB symmetry can be formulated as a 
requirement that physical results should be
invariant under this transformation.

\begin{sloppypar}
Let us now apply the AB transformation twice. 
Then, by analogy with the derivation of Eq. (\ref{54}),
we get 
\begin{equation}
a(n_1n_2nk)^*\rightarrow \mp a(n_1n_2nk)^*\quad
a(n_1n_2nk)\rightarrow \mp a(n_1n_2nk)
\label{79}
\end{equation} 
for the normal and broken spin-statistic connections,
respectively. Therefore, as a consequence of the 
spin-statistics theorem, any particle
(with the integer or half-integer spin) has the
AB$^2$ parity equal to $-1$. Therefore in GFQT any 
interaction can involve only an even number of
creation and annihilation operators. In particular,
this is additional demonstration of the fact that in
GFQT the existence of neutral elementary particles
is incompatible with the spin-statistics theorem.
\end{sloppypar}

\section{Discussion}

In the present paper we discuss the description of 
free elementary particles in a
quantum theory based on a Galois field (GFQT). 
As noted in Sect. \ref{S1}, GFQT does not contain
infinities at all and all operators are 
automatically well defined. In my discussions with
physicists, some of them commented this fact as 
follows. This is the approach where a cutoff
(the characteristic $p$ of the Galois field) is
introduced from the beginning and for this reason
there is nothing strange in the fact that
the theory does not have infinities. It has a 
large number $p$ instead and this number can be
practically treated as infinite. 

However, the difference between Galois fields 
and usual numbers is not only that the former are 
finite and the latter are infinite. If the set of 
usual numbers is visualized as a straight line from 
$-\infty$ to $+\infty$ then the simplest Galois field
can be visualized not as a segment of this line but
as a circle (see Ref. \cite{hep} for a detailed discussion). 
This reflects the fact that in Galois fields the rules of 
arythmetic are different and, as a result, GFQT has 
many unusual features which have no
analog in the standard theory. 

The Dirac vacuum energy problem discussed in Sect. 
\ref{S6} is a good illustration of this point. 
In the standard theory the result that the vacuum energy is zero
is a consequence of the requirement that all
operators should be written in the normal form. 
This requirement does not follow from the theory and is imposed with
the only purpose to obtain the correct result for the vacuum energy
and other observables. Without this requirement, the vacuum energy
in the standard theory is infinite. Therefore 
if GFQT is treated simply as a theory with a cutoff
$p$, one would expect the vacuum energy 
to be of order $p$. However, since the rules
of arithmetic in Galois fields are different from the
standard ones, the result for the vacuum energy is
exactly zero. The consideration of the vacuum energy 
also poses the following very interesting problem.
The result is based on the prescription of Sect. 
\ref{S5} for separating physical and nonphysical 
states. With such a prescription the vacuum energy is
zero only for particles with the spin 1/2. Is this an
indication that only such particles can be elementary
or the prescription (although it seems very
natural) should be changed?

\begin{sloppypar}
The original motivation for investigating GFQT was
as follows. Let us take standard QED in
dS or AdS space, write the Hamiltonian and other
operators in angular momentum basis and 
replace standard irreducible representations (IRs)
for the electron, positron and
photon by corresponding modular IRs. One might treat 
this motivation as an attempt to substantiate
standard momentum regularizations (e.g. the
Pauli-Villars regularization) at momenta $p/R$
(where $R$ is the radius of the Universe).
In other terms this might be treated as introducing
fundamental length of order $R/p$. We now discuss
reasons explaining why this naive attempt fails.
\end{sloppypar}

The main result of the present paper is that {\it in GFQT
the existence of antiparticles follows from the fact
that any Galois field is finite. Moreover, one can say that
the very existence of antiparticles is a strong indication that nature 
is described rather by a finite field (or at least a field with
a nonzero characteristic) than by complex numbers.} We believe that this 
result is not only very important but also extremely 
simple and beautiful. A mathematical consideration
of modular IRs is given in Sects. \ref{S2} 
and \ref{S3} while a simple explanation of the above 
result is as follows. 

In the standard theory a particle is described by a
positive energy IR where the energy has the spectrum
in the range $[mass,\infty)$. At the same time, the
corresponding antiparticle is associated with a
negative energy representation where the energy has 
the spectrum in the range $(-\infty,-mass]$. 
Consider now the construction of modular IR for some
particle. We again start from the rest state (where
energy=mass) and gradually construct states with
higher and higher energies. However, since a Galois field
is not a straight line but rather a circle, sooner or 
later we will arrive at the point where energy=-mass. 

In QFT the fact that a particle and its antiparticle have the
same masses and spins but opposite charges follows from the CPT theorem,
which is a consequence of locality. A question arises what happens if
locality is only an approximation: in that case the equality
of masses, spins etc. is exact or approximate?
Consider a simple model when electromagnetic and weak interactions are
absent. Then the fact that the proton and the neutron have the same
masses and spins has nothing to do with locality; it is only a
consequence of the fact that the proton and the neutron belong to the
same isotopic multiplet. In other words, they are simply different
states of the same object - the nucleon. We see, that in GFQT the
situation is analogous. The
fact that a particle and its antiparticle have the
same masses and spins but opposite charges has nothing to do with locality
or nonlocality and is simply a consequence of the fact that they are
different states of the same object since they belong to the same IR.

In standard theory, a particle and its antiparticle
are combined together by a local covariant equation 
(e.g. the Dirac equation). We see that in GFQT
the idea of the Dirac equation is implemented without
assuming locality but already at the level of IRs.
This automatically explains the existence of 
antiparticles, shows that a particle cannot exist 
by itself without its antiparticle and that a particle
and its antiparticle are necessarily {\it different} 
states of the same object. In particular, there are
no elementary particles which in the standard theory 
are called neutral. Note also that, as shown in
Ref. \cite{JPA}, even in {\it standard} theory
with the invariance group SO(1,4) the only 
possible interpretation of IRs is that they describe
a particle and its antiparticle simultaneously.

One might conclude that since in GFQT the
photon cannot be elementary, this theory cannot be
realistic and does not deserve attention. 
We believe however, that the nonexistence of 
neutral elementary 
particles in GFQT shows that the photon, the graviton
and other neutral particles should be considered on a
deeper level. For example, several authors considered
a model where the photon is a composite state of Dirac
singletons \cite{FF}. 

The nonexistence of neutral elementary particles in GFQT
is discussed in detail in Sect. \ref{S7}. A possible
elementary explanation is as follows. In the standard
theory, elementary particles having antiparticles are
described by complex fields while neutral elementary
particles --- by real ones. There also exists the
quaternionic version of the standard theory. In GFQT
however, the quaternionic version cannot exist and there 
is no possibility to choose between real and complex fields.
If $p=3\,(mod\, 4)$ then 
analogs of real fields can exist only if the spin-statistics
connection is broken while the spin-statistics theorem can 
be treated as a requirement that standard quantum theory should be 
based on complex numbers. This requirement excludes the
existence of neutral elementary particles.

The results of Sect. \ref{S5} show that in GFQT
there exists the following dilemma. If the notion
of particles and antiparticles is valid at all
energies then only particles with the half-integer
spin (in the usual units) can be elementary.
Therefore supersymmetry is possible only if the
notion of particles and particles is not valid
at asymptotically large energies. This problem
requires further study. 

{\it Acknowledgements:} The author is grateful to F. Coester,
H. Doughty, C. Hayzelden, M. Planat, W. Polyzou and M. Saniga 
for useful discussions.

\end{document}